\documentclass{PoS}
\pdfoutput=1
\usepackage{amsmath,bbm}

\DeclareMathOperator\diag{diag}
\DeclareMathOperator\ind{ind}
\DeclareMathOperator\re{Re}
\DeclareMathOperator\SU{SU}
\DeclareMathOperator\Sp{Sp}
\DeclareMathOperator\tr{tr}
\DeclareMathOperator\U{U}
\newcommand\1{\mathbbm 1}
\newcommand\eps{\varepsilon}
\newcommand\ev[1]{\left\langle#1\right\rangle}
\newcommand\fh{\tilde f}
\newcommand\vh{\tilde v}
\renewcommand{\H}{\textbf{H}}
\newcommand{\I}{\textbf{I}}
\renewcommand{\L}{\textbf{L}}
\newcommand\mL{\mathcal L}
\newcommand\rsv{\rho_\text{sv}}

\title{Singular values of the Dirac operator at nonzero
  density\thanks{Supported by the Alexander von Humboldt Foundation,
    DFG, and JSPS.}} 

\ShortTitle{Singular values of the Dirac operator at nonzero density}

\author{Takuya Kanazawa\\
  Department of Physics, University of Regensburg, 93040 Regensburg,
  Germany\thanks{Affiliation until March 2012.}}
\author{\speaker{Tilo Wettig}\\
  Department of Physics, University of Regensburg, 93040 Regensburg,
  Germany\\
  E-mail: \email{tilo.wettig@ur.de}}
\author{Naoki Yamamoto\\
  Yukawa Institute for Theoretical Physics,
  Kyoto University, Kyoto 606-8502, Japan\\
  Institute for Nuclear Theory, University of Washington,
  Seattle, Washington 98195-1550, USA\\
  Department of Physics, University of Maryland, College Park, MD
  20742-4111, USA\\
  E-mail: \email{nyama@umd.edu}}

\abstract{At nonzero density the eigenvalues of the Dirac operator
  move into the complex plane, while its singular values remain real
  and nonnegative. In QCD-like theories, the singular-value spectrum
  carries information on the diquark (or pionic) condensate. We have
  constructed low-energy effective theories in different density
  regimes and derived a number of exact results for the Dirac singular
  values, including Banks-Casher-type relations for the diquark (or
  pionic) condensate, Smilga-Stern-type relations for the slope of the
  singular-value density, and Leutwyler-Smilga-type sum rules for the
  inverse singular values. We also present a rigorous index theorem
  for non-Hermitian Dirac operators.}

\FullConference{The 30th International Symposium on Lattice Field Theory\\
  June 24 -- 29,  2012\\
  Cairns, Australia}

\begin{document}

\section{Introduction}

We give a summary of \cite{Kanazawa:2011tt}, which is very long and
sometimes quite technical.  Here we present the main ideas and most
important results.  We work in Euclidean space and in the chiral
limit.

\section{Dirac eigenvalues and singular values}

In the presence of a chemical potential $\mu$, the massless Dirac
operator in QCD and QCD-like theories is given by
\begin{align}
  D(\mu)=\gamma_\nu D_\nu+\mu\gamma_4
  =\begin{pmatrix}0&D_L\\D_R&0\end{pmatrix}
\end{align}
with $D_\nu=\partial_\nu+iA_\nu$, where $A_\nu=A_\nu^a\tau_a/2$ is the
gauge field and the $\tau_a$ are the generators of the gauge group in
the representation in which the fermions transform.  The eigenvalue
equation and the equation for the singular values are
\begin{align}
  D(\mu)\psi_n=\lambda_n\psi_n\qquad\text{and}\qquad
  D^\dagger D\phi_n=\xi_n^2\phi_n\,,
\end{align}
where the argument $\mu$ is understood if it is not shown explicitly.
For $\mu=0$ we have the trivial relation $\xi_n=|\lambda_n|$.  For
nonzero (and real) $\mu$ the eigenvalues move into the complex plane,
while the singular values are always real and nonnegative.
Eigenvalues and singular values are unrelated for $\mu\ne0$ and live
on different physical scales.  The scale for the eigenvalues is set by
the chiral condensate at low density \cite{Banks:1979yr} and by the
BCS gap $\Delta$ at high density \cite{Yamamoto:2009ey}, while the
scale for the singular values in QCD-like theories is set by the
diquark condensate at all densities, see Sec.~\ref{sec:bc}.

The operators $D(\mu)$ and $D^\dagger D$ have the same zero modes, and
the same is true for the operators $D(-\mu)$ and $DD^\dagger$.  For a
non-Hermitian Dirac operator the index theorem takes the form
\begin{align}
  \label{eq:ind}
  \frac1{32\pi^2}\int d^4x\,F\tilde F
  =\frac12\big[\ind D(\mu)+\ind D(\mu)^\dagger\big],
\end{align}
where $\ind D(\mu)=\dim\ker D_R-\dim \ker D_L$.  We generically have
$\ind D(\mu)=\ind D(\mu)^\dagger$ unless the gauge field is
fine-tuned.  The proof of \eqref{eq:ind} and further properties of the
eigenvalues and singular values are given in \cite{Kanazawa:2011tt}.

\section{Banks-Casher-type relations for the singular values}
\label{sec:bc}

For the derivation of the results in this section it is important that
the fermionic measure (including source terms) in the partition
function is positive definite, see the discussion in
\cite[App.~C]{Kanazawa:2011tt}.  We restrict ourselves to such cases,
which in particular implies an even number $N_f$ of flavors (except
for adjoint QCD, where $N_f$ could also be odd).

\subsection{Two-color QCD}

In two-color QCD the Dirac operator satisfies an anti-unitary
symmetry, $[C\tau_2K,iD]=0$ with $C=i\gamma_4\gamma_2$ the charge
conjugation matrix and $K$ the operator of complex conjugation.
Adding a source term $\tfrac12j\psi^TC\gamma_5\tau_2I\psi+\text{c.c.}$
with
\begin{align}
  I=\begin{pmatrix}
    0 & -\1_{N_f/2}\\
    \1_{N_f/2} & 0\\
  \end{pmatrix}
\end{align}
to the fermionic Lagrangian and using the Nambu-Gor'kov formalism one
can show that the partition function takes the form
\begin{align}
  \label{eq:Zj}
  Z(j)=\ev{{\det}^{N_f/2}(D^\dagger D+j^2)}_\text{YM}
  =\Big\langle\prod_n(\xi_n^2+j^2)^{N_f/2}\Big\rangle_\text{YM}\,,
\end{align}
where YM stands for $N_f=0$ in the average over gauge fields.  Taking
the logarithmic derivative w.r.t.\ $j$, followed by the limits
four-volume $V_4\to\infty$ and $j\to0^+$ (in this order), then yields
a relation between the scalar diquark condensate and the density
$\rsv(\xi)$ of the singular values at the origin,
\begin{align}
  \ev{\psi^TC\gamma_5\tau_2I\psi}=\frac{N_f}2\pi\rsv(0)\,,
\end{align}
see also \cite{Fukushima:2008su}.  Some subtleties of this result and
its derivation are discussed in \cite{Kanazawa:2011tt}.

\subsection{QCD with isospin chemical potential}

We now consider QCD with $N_c\ge2$ colors, two flavors, and nonzero
isospin chemical potential $\mu_I = 2\mu$, i.e., we have $D(\mu)$ for
the up quark and $D(-\mu)$ for the down quark.  Adding a source term
$j(d_L^\dagger u_R-u_L^\dagger d_R)+\text{c.c.}$ to the Lagrangian and
using $D(\mu)^\dagger=-D(-\mu)$ we find
\begin{align}
  Z(j)=\ev{\det(D^\dagger D+j^2)}_\text{YM}\,,
\end{align}
which agrees with \eqref{eq:Zj} for $N_f=2$.  After differentiating
$\ln Z(j)$ w.r.t.\ $j$ we obtain a relation between the pionic
condensate and $\rsv$, i.e.,
\begin{align}
  \ev{\bar u\gamma_5 d-\bar d\gamma_5 u}=\pi\rsv(0)\,.
\end{align}

\subsection{Adjoint QCD}

Turning to QCD with adjoint fermions and any number of colors, we add
a source term $\tfrac12j\psi^TC\gamma_5\psi+\text{c.c.}$ and again
obtain \eqref{eq:Zj}.  An analogous calculation then yields
\begin{align}
  \ev{\psi^TC\gamma_5\psi}=\frac{N_f}2\pi\rsv(0)\,.
\end{align}

\section{Low-energy effective theories with diquark sources}
\label{sec:eff}

From now on we focus on two-color QCD and distinguish three density
regimes that differ in their pattern of chiral symmetry breaking and
in the number of Nambu-Goldstone (NG) modes.

\subsection{Low density}

This regime was analyzed in great detail in
\cite{Kogut:1999iv,Kogut:2000ek}.  Here one starts from the symmetry
breaking pattern $\SU(2N_f)\to\Sp(2N_f)$ at zero density and treats
$\mu$ and the diquark source as a small perturbation.  There are $N_f
(2N_f-1)-1$ NG modes, parametrized by a field $\Sigma=U\Sigma_dU^T$
with $\Sigma_d=\diag(I,-I)$ and $U=\exp(i \pi^a T^a/2F)$, where the
$T^a$ are the generators of the coset space $\SU(2N_f)/\Sp(2N_f)$ and
$F$ is a low-energy constant (LEC).  The leading-order effective
Lagrangian in the chiral limit is
\begin{align}
  \label{eq:Lefflow}
  \mL_\text{eff}^\L=\frac{F^2}2\tr(\nabla_\nu\Sigma\nabla_\nu\Sigma^\dagger)
  -\Phi_{\L}\re\tr(\bar J\Sigma)
\end{align}
with \allowdisplaybreaks[4]
\begin{align}
  \nabla_\nu\Sigma&=\partial_\nu\Sigma-\mu\delta_{\nu0}
  (B\Sigma+\Sigma B)\,,&
  \nabla_\nu\Sigma^\dagger&=\partial_\nu\Sigma^\dagger
  +\mu\delta_{\nu0}(\Sigma^\dagger B+B\Sigma^\dagger)\,,\\
  B&=\begin{pmatrix}\1_{N_f} & 0 \\ 0 & -\1_{N_f}\end{pmatrix},
  & \bar J&=\begin{pmatrix} J_L & 0 \\ 0 & -J_R^\dagger \end{pmatrix}.
  \label{eq:Jbar}
\end{align}
Here, $J_{L/R}$ are antisymmetric complex matrices of dimension $N_f$
in flavor space, and $\Phi_{\L}$ is another LEC equal to the diquark
condensate per flavor and handedness at $\mu=0$ and without sources,
\begin{align}
  \label{eq:PhiL}
  \Phi_{\L}=\frac1{N_f}\left|\ev{\psi^T_iC\tau_2I\psi_i}
  \right|_{\bar J=0,\,\mu=0}\qquad(i=L,R)\,.
\end{align}
For $J_R=-J_L=jI$ there are two types of NG modes \cite{Kogut:2000ek},
\begin{subequations}
  \label{eq:GOR_low}
  \begin{align}
    \label{eq:GOR_low_a}
    \text{type 1: } &&& \text{mass}=\sqrt{j\Phi_\L/F^2} &&
    (N_f^2-N_f-1\text{ modes})\,,\\
    \label{eq:GOR_low_b}
    \text{type 2: } &&& \text{mass}=\sqrt{j\Phi_\L/F^2+(2\mu)^2} &&
    (N_f^2\text{ modes})\,.
  \end{align}
\end{subequations}
The type-2 modes become massive for $\mu\ne0$, while the type-1 modes
stay massless for $j\to0$.

\subsection{Intermediate density}
\label{sec:int}

At intermediate density $\mu$ can no longer be treated as a small
perturbation and breaks the original $\SU(2N_f)$ symmetry to
$\SU(N_f)_L\times\SU(N_f)_R\times\U(1)_B$.  A diquark condensate then
breaks this symmetry to $\Sp(N_f)_L\times\Sp(N_f)_R$.
The corresponding NG modes are $\Sigma_L,\Sigma_R\in\SU(N_f)/\Sp(N_f)$
and $V\in\U(1)_B$, and the total number of NG modes in this regime is
$N_f(N_f-1)-1$.  The $\U(1)_A$ symmetry is broken explicitly by the
anomaly.  The effective Lagrangian is
\begin{align}
  \mL_\text{eff}^\I=\left\{ 
    \begin{array}{l}      
      N_ff_0^2\left[|\partial_0 V|^2
        +v_0^2|\partial_i V|^2 \right] +\frac{f^2}{2}\tr\left[|\partial_0
        \Sigma_L |^2 +v^2 |\partial_i \Sigma_L |^2 
        + (L \leftrightarrow R)\right]\\
      \quad-\Phi_{\I} \re\left[V\tr (J_L\Sigma_L-J_R\Sigma_R)\right]
      \hfill\text{for }N_f\ge4\,,\\[3mm]
      2f_0^2\left[|\partial_0 V|^2         
        +v_0^2|\partial_i V|^2 \right]+2\Phi_{\I}\re[(j_L-j_R)V]
      \hfill\text{for }N_f=2\,,
    \end{array}
    \right.
  \label{eq:int}  
\end{align}
where $f_0,f$ and $v_0,v$ are LECs corresponding to the decay
constants and velocities of the NG modes, respectively, and
$\Phi_{\I}$ is an LEC similar to \eqref{eq:PhiL} but with $\mu\ne0$.
All LECs depend on $\mu$.  The masses of the NG modes are given by
\begin{align}
  \label{eq:GOR_int}
  m_A=\sqrt{j\Phi_\I/f_A^2}\quad\text{for }A=0,\ldots,N_f(N_f-1)-2\,,
\end{align}
where $f_A=f$ for $A\ge1$.  Note that in this regime all NG modes are
massless in the $j\to0$ limit.

\subsection{High density}

At very high density the $\U(1)_A$ anomaly is suppressed due to the
screening of instantons \cite{Schafer:2002ty,Schafer:2002yy}.  The
$\U(1)_A$ symmetry of the action is no longer broken explicitly by the
anomaly but spontaneously by the diquark condensate.  Therefore the
symmetry-breaking pattern is now \cite{Kanazawa:2009ks}
\begin{align}
  \label{eq:SBP_high}
  \SU(N_f)_L\times\SU(N_f)_R\times\U(1)_B\times\U(1)_A
  \to\Sp(N_f)_L\times\Sp(N_f)_R\,.
\end{align}
The NG modes are the same as at intermediate density, except that
there is an additional NG mode $A\in\U(1)_A$ which can be considered
to be the $\eta'$ whose mass has become small.  Hence the total number
of NG modes in this regime is $N_f(N_f-1)$.  The effective Lagrangian
is 
\begin{align}
  \label{eq:lagrangian_high}
  \mL_\text{eff}^\H=\left\{
    \begin{array}{l}
      \Big[\frac{N_f\fh_0^2}{2}\big(|\partial_0 L|^2
      +\vh_0^2|\partial_i L|^2 \big)+\frac{\fh^2}{2}\tr\big(
      |\partial_0 \Sigma_L |^2+\vh^2 |\partial_i \Sigma_L |^2 \big)
      + (L \leftrightarrow R)\Big]\\
      \quad-\Phi_{\H} \re\tr(J_LL\Sigma_L-J_RR\Sigma_R)
      -\frac{2\fh_0^2}{N_f}m_\text{inst}^2\re\,(L^\dagger R)^{N_f/2}
      \hfill\text{for }N_f\ge4\,,\\[3mm]
      \fh_0^2\left[|\partial_0 L|^2+\vh_0^2|\partial_i L|^2 
        + (L \leftrightarrow R)\right] + 2\Phi_{\H} \re(j_LL-j_RR) 
      -\fh_0^2m_\text{inst}^2 \re(L^\dagger R)
      \;\;\;\text{for }N_f=2\,,
    \end{array}
    \right.
\end{align}
where $L=A^\dagger V$ and $R=AV$.  Similar to Sec.~\ref{sec:int} we
have LECs $\fh_0,\fh$ and $\vh_0,\vh$ as well as $\Phi_{\H}$, all of
which depend on $\mu$.  The term involving $m_\text{inst}$ in
\eqref{eq:lagrangian_high} corresponds to the single-instanton
contribution to the $\eta'$ mass, with $m_\text{inst}\to0$ as
$\mu\to\infty$, see \cite{Kanazawa:2011tt} for a detailed discussion.
The masses of the NG modes are given by
\begin{subequations}
  \label{eq:GOR_high}
  \begin{align}
    \label{eq:GOR_high_a}
    \text{type 1: } &&& m_A=\sqrt{j\Phi_{\H}/\fh_A^2} &&
    (N_f^2-N_f-1\text{ modes})\,,\\
    \label{eq:GOR_high_b}
    \text{type 2: } &&&
    m_{\eta'}=\sqrt{j\Phi_\H/\fh_0^2+m_\text{inst}^2}
    && (1\text{ mode})\,.
  \end{align}
\end{subequations}
The type-1 modes stay massless for $j\to0$, while the $\eta'$ 
becomes massive as $\mu$ is lowered.

The effective theory at intermediate density can be obtained from the
effective theory at low density (or high density) by integrating out
the NG modes that become massive as the density is increased (or
decreased).  In this way the LECs of the three different regimes can
be matched, see \cite{Kanazawa:2011tt} for details.  The domains of
validity of the three effective theories, and their overlaps, are also
discussed in \cite{Kanazawa:2011tt}.

\section{Smilga-Stern-type relations for the singular values}

Following the approach of Smilga and Stern for QCD at zero density
\cite{Smilga:1993in}, we can compute the slope of the singular-value
density at the origin.  For technical reasons we start at infinite
density.

\subsection{Infinite density}

We now add a more general source term
$\tfrac12\psi^TC\gamma_5\tau_2J\psi+\text{c.c.}$ to the fermionic
Lagrangian, where $J=J_R=-J_L$ is again an antisymmetric $N_f\times
N_f$ matrix, which we can decompose as
\begin{align}
  \label{eq:J}
  J=I\sum_{A=0}^Nj_At^A=jI+I\sum_{a=1}^Nj_at^a
  \quad\text{with}\quad N=\frac12N_f(N_f-1)-1\,.
\end{align}
Here, the $t^A$ are the generators of $\U(N_f)/\Sp(N_f)$ and the $j_A$
are real parameters with $j_0=j\sqrt{N_f}$.  We can then show the
partition function of two-color QCD is given by
\begin{align}
  Z(J)=\ev{{\det}^{1/2}(D^\dagger D+J^\dagger J)}_\text{YM}
  =\Big\langle\prod_n{\det}^{1/2}(\xi_n^2+J^\dagger
  J)\Big\rangle_\text{YM}\,.
\end{align}
For $N_f\ge4$ we define the scalar susceptibility
\begin{align}
  \label{eq:Kab}
  K_{ab}(j)=\lim_{V_4\to\infty}\frac1{V_4}\partial_{j_a}\partial_{j_b}
  \ln Z(J)\Big|_{\text{all }j_a=0}\,,
\end{align}
for which we obtain after some algebra
\begin{align}
  K_{ab}(j) &= \delta_{ab}\lim_{V_4 \rightarrow \infty} \frac{1}{V_4}
  \bigg\langle \sum_n\frac{\xi_n^2-j^2}{(\xi_n^2+j^2)^2}
  \bigg\rangle_j = \delta_{ab}\int_0^{\infty} d\xi \,\rsv(\xi;j)
  \frac{\xi^2-j^2}{(\xi^2+j^2)^2}
  \sim\delta_{ab}\rsv'(0)\ln\frac{\tilde\Lambda}j\,.
  \label{eq:KQCD}
\end{align}
In the last step we cut off the integral at $\xi=\tilde\Lambda$
and extracted the part that diverges for $j\to0$.  On the low-energy
effective theory side we start from \eqref{eq:lagrangian_high}, where
at infinite density we can set $m_\text{inst}=0$.  Performing a
one-loop calculation with a momentum cutoff $\tilde\Lambda$ we obtain
for $j\to0$
\begin{equation}
  \label{eq:Keff2}
  K_{ab}(j)\sim\delta_{ab}\left[\frac{(N_f-4)(N_f+2)}{2N_f \fh^4}
    +\frac{2}{N_f \fh_0^2 \fh^2} \right]
  \frac{\Phi_\H^2}{16\pi^2} \ln\left(\frac{\tilde\Lambda}{j}\right).
\end{equation}
Matching the divergences of \eqref{eq:KQCD} and \eqref{eq:Keff2} for
$j\to0$ yields the slope of $\rsv$ at the origin,
\begin{align}
  \label{eq:slope_high}
  \rsv'(0)=
  \left[\frac{(N_f-4)(N_f+2)}{2N_f \fh^4}+\frac{2}{N_f \fh_0^2 \fh^2}
  \right]
  \frac{\Phi_\H^2}{16\pi^2}\,.
\end{align}
This method does not work for $N_f=2$, but we expect, based on
experience from partially quenched chiral perturbation theory
\cite{Toublan:1999hi}, that \eqref{eq:slope_high} is also valid for
$N_f=2$.

\subsection{Intermediate density}

At intermediate density the calculation on the low-energy effective
theory side starts from \eqref{eq:int} and proceeds in a similar way
to yield
\begin{align}
  \label{eq:slope_int}
  \rsv'(0)=
  \left[\frac{(N_f-4)(N_f+2)}{2N_f f^4}+\frac{1}{N_f f_0^2 f^2}
  \right]
  \frac{\Phi_\I^2}{16\pi^2}\,.  
\end{align}

\subsection{Zero density}

At strictly zero density we start from \eqref{eq:Lefflow} with $\mu=0$
and obtain in a similar way
\begin{align}
  \label{eq:slope_zero}
  \rsv'(0)=\frac{(N_f-2)(N_f+1)}{N_f F^4}\frac{\Phi_\L^2}{16\pi^2}\,.
\end{align}

\subsection{Relation between the three results}

At first sight it does not seem possible to interpolate smoothly
between the three results for $\rsv'(0)$ at zero, intermediate, and
infinite density.  This puzzle can be understood by analogy with the
interpolation between SU(2) and SU(3) chiral perturbation theory,
where the strange quark mass plays the role of the symmetry-breaking
parameter.  In our case this parameter is $\mu$.  We find (see
\cite{Kanazawa:2011tt} for details) that at low density the slope is
given by \eqref{eq:slope_int} for $\xi\ll\mu^2/\Lambda$ and by
\eqref{eq:slope_zero} for $\mu^2/\Lambda\ll\xi\ll\Lambda$, where
$\Lambda\sim F\sim\Phi_\I^{1/3}$.  At high density the slope is given
by \eqref{eq:slope_int} for $\xi\ll{gm_{\eta'}^2}/\Delta$ and by
\eqref{eq:slope_high} for ${gm_{\eta'}^2}/\Delta\ll\xi\ll g\Delta$,
where $g$ is the coupling constant.

\section{Finite-volume analysis}

In a finite volume $V_4=L^4$, the three low-energy effective theories
constructed in Sec.~\ref{sec:eff} have $\eps$-regimes in which the
theory becomes zero-dimensional.
The corresponding condition is
\begin{align}
  \label{eq:eps}
  \frac1{m_\ell}\ll L\ll\frac1{m_\text{NG}}\,,
\end{align}
where $m_\ell$ is the mass scale of the lightest non-NG particle (see
\cite[Sec.~5.5]{Kanazawa:2011tt} for actual values) and $m_\text{NG}$
is the mass scale of the NG particles, see Eqs.~\eqref{eq:GOR_low},
\eqref{eq:GOR_int}, and \eqref{eq:GOR_high}.  In these $\eps$-regimes
we have derive Leutwyler-Smilga-type sum rules \cite{Leutwyler:1992yt}
for the inverse singular values and random matrix theories from which
microscopic singular-value correlation functions can be derived
\cite{Kanazawa:2011tt}.

\section{Summary}

We have derived a number of exact results for the singular-value
spectrum in QCD-like theories without a sign problem.  Our results
could be used in lattice simulations to obtain the diquark condensate
at any density, allowing for a numerical test of the conjectured
BEC-BCS crossover.

\bibliographystyle{JBJHEP}
\bibliography{myref}

\end{document}